\documentclass[showpacs,showkeys,preprintnumbers,amsmath,amssymb,twocolumn,prl,floatfix]
 {revtex4}

\usepackage{amssymb,amsfonts}
\usepackage{amsmath}
\usepackage{hyperref}
\usepackage{graphicx}
\usepackage[usenames,dvipsnames]{xcolor}
\begin{document}

\title{Dynamical model of the kinesin protein motor }

\author{Alexander I Nesterov}%
  \email{nesterov@cencar.udg.mx}
\affiliation{Departamento de F{\'\i}sica, CUCEI, Universidad de Guadalajara,
Av. Revoluci\'on 1500, Guadalajara, CP 44420, Jalisco, M\'exico}

\author{Gennady P  Berman}
 \email{gpb@lanl.gov}
\affiliation{Biology Division, B-11, Los Alamos National Laboratory, and 
the New Mexico Consortium,  Los Alamos, NM 87544, USA}

\author{M\'onica F Ram\'irez }
 \email{monica.felipa@gmail.com}
\affiliation{ Departamento de F{\'\i}sica, CUCEI, Universidad de 
Guadalajara,
Av. Revoluci\'on 1500, Guadalajara, CP 44420, Jalisco, M\'exico}
 
\date{\today}

\begin{abstract}
We model and simulate the stepping dynamics of the kinesin motor including electric and mechanical forces, environmental noise, and the complicated   potentials produced by tracking and neighboring 
protofilaments. Our dynamical model supports the hand-over-hand 
mechanism of the kinesin stepping. Our theoretical predictions and 
numerical simulations include the off-axis displacements of the kinesin heads 
while the steps are performed. The results obtained are in a good agreement 
with recent experiments on the kinesin dynamics.

\end{abstract}



\preprint{LA-UR-16-22931 }

\maketitle

Microtubules (MTs) are cylindrically shaped cytoskeletal biopolymers. They are found in eukaryotic cells and are formed by the polymerization of a heterodimer of two globular proteins, $\alpha$ and $\beta$ tubulin. Under suitable conditions, tubulin heterodimers assemble into linear protofilaments (PFs), and MTs are realized as hollow cylinders typically formed by $13$ parallel PFs covering the wall of MT. The outer diameter of a MT is about 25 nm, and the inner diameter is about 15 nm. In the ground state the MT has a permanent dipole momentum along the MT. This results in the uniform electric field along the MT \cite{Amos,TBHM,BSJ,NRB}.

Kinesin and related motor proteins convert the chemical energy of  ATP  hydrolysis to move themselves along the PF. Conventional kinesin (kinesin-1) moves progressively from the ``$-$" end to the ``$+$" end of the MT. It can make 100 or more steps before dissociating from the MT. The observed maximum propagation velocity of  kinesin in vivo is  between $600\, \rm nm/s$ and $2000 \, \rm nm/s $, and it depends on various chemical and physical factors, especially on the concentration of ATP \cite{BTU,BTS,BSV,VRS,MMM}.

 Currently, a commonly accepted mechanism of the kinesin stepping is the {\em hand-over-hand mechanism} \cite{HVR,HHJ,HD,SHJ,AFB,Yild04,Schaap2011}. The hand-over-hand model assumes that the kinesin uses both of its motors by alternately stepping along a single PF. The hand-over-hand model predicts that, for each ATP hydrolysis cycle, the rear head moves around the front (immobile) head performing the displacement $2l_0$, where $l_0$ is the distance between the heads attached to the MT \cite{Yild04,Schaap2011}.  

One challenge is to explain recent experiments on the kinesin dynamics which demonstrated a strong off-axis displacement of the kinesin head while the step is performed, the structure of  sub-steps, and other details of the kinesin dynamics \cite{MDO,HRY}. Prior models of the kinesin stepping cannot explain these results, since in these models the free (detached) kinesin head only moves in the plane orthogonal to the MT surface \cite{Block2007,TIK,SQG,CCA}. 

In our model, the step-like motion is due to time-varying charge 
distributions, following by the ATP hydrolysis cycle. The uniform electric 
field on the surface of the MT has two functions: to push the trailing head 
forward to the next docking site and to load the coiled spring. The 
asymmetric torsional energetic barriers in the coiled coil region  appear 
because of chirality of the coiled-coils \cite{AFB,BSG}. 

Thus, when the 
trailing head steps forward, the neck coiled-coil is overwound relative to the 
relaxed state. At the next step, when the other head moves, the coiled 
spring is underwound, and the system is returned to its initial state, with the 
both heads shifted by $2l_0$ and being attached to the MT, and the neck coiled-coil relaxed.

{\em Modeling the Stepping Dynamics.} -- The kinesin-MT binding-interface is dominated by ionic interactions. Since initially both heads are positively charged, the interaction of the heads with the negatively charged tubulin heterodimers provides a stable (locked)  state of the kinesin \cite{Woehlke97,GGZ}.

\begin{figure}[tbh]
  \begin{center}
 \scalebox{0.175}{\includegraphics{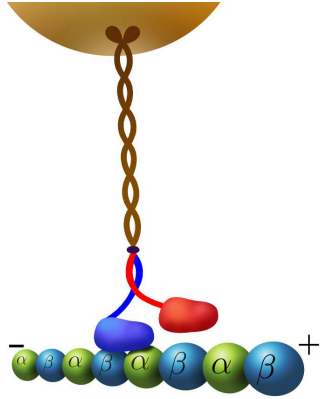}}
 \end{center}
 \caption{(Color online) Kinesin
walking along a MT. 
 \label{Kinesin_1a}}
  \end{figure}

Before the first step begins, both heads of the kinesin are attached to 
neighboring $\beta$-tubulins. When an ATP binds into the rear attached 
head, the ATP hydrolysis cycle begins, and a general reorganization of the 
motor domain occurs. This leads to local charge redistribution: the 
kinesin head becomes negatively charged and electrostatically unstable.  As 
a result, the trailing head is unlocked and begins to move in the electric field 
around the leading head, until reaching the closest docking site 
($\beta$-tubulin subunit). Due to the continuing redistribution of the 
charge, the trailing head becomes positively charged and attached to the 
pocket located at the $\beta$-tubulin subunit. Thus, every kinesin step 
includes docking of the leading head  to the 
$\beta$-tubulin followed by rotation of the trailing head.  (See Fig. 
\ref{Kinesin_1a}.)  

Since the mass of the kinesin head is very small ($m \approx 6.3\cdot10^{-19}\rm g$) \cite{AGL}, the kinesin dynamics can be described by an over-damped Langevin equation:
\begin{align}
	\gamma \frac{d \mathbf r}{d t} = - \nabla V(\mathbf r, t) + {\mathbf F}_{ex} (\mathbf r ,t) + {\boldsymbol \xi} (t),
	\label{EqL}
\end{align}
where $\gamma$ is the damping coefficient, $V$ is the potential energy of the trailing head interacting with the surrounding electric field, and $\mathbf F_{ex}$ describes external forces. The environment is simulated by the thermal noise, ${\boldsymbol \xi}  (t)$, with the properties:  $\langle \xi_i (t) \rangle =0$ and  $\langle \xi_i (t) \xi_j (t') \rangle =2\gamma \rm  k_BT \delta_{ij} \delta (t -t')$, where $T$ is the temperature. 

The available experimental data on the behavior of the kinesin motor proteins show that the drag  (stall) forces exerted on them in living cells are: $F\sim(4 \div 7 )\, \rm pN$ \cite{SBS,HHG,LLG}. The maximum velocity of the kinesin in vivo can be estimated as: $\langle v\rangle =(1800 \div 2000)\, \rm nm/s$ \cite{BTS}.  Using the relation: $F=\gamma \langle v\rangle $, one can estimate the damping coefficient as: $\gamma \sim (2\div 3.5 )\cdot 10^{-3} \, \rm pN \cdot s/nm$.

\begin{figure}[tbh]
  \begin{center}
    \scalebox{0.36}{\includegraphics{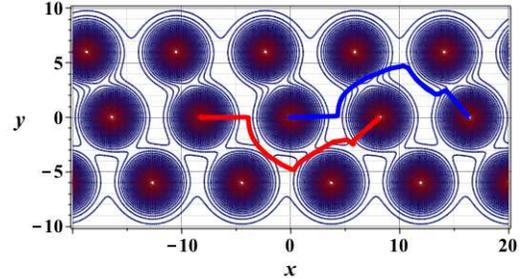}}
\end{center}
 \caption{(Color online) Docking potential landscape. The red and blue curves show the trajectories of the right and the left heads, respectively. The stiffness of the neck linker is: $k = 1.75\, \rm pN /nm$, $\tau_\alpha = 0.5\, \rm pN\cdot nm \cdot rad^{-1}$. 
 \label{DP2a}}
  \end{figure}

 We choose the $x$-axis along the PF, and we assume that the kinesin moves in the  $(x,y)$-plane, in the positive  $x$-direction. The potential energy of the system can be separated into two parts: 
$V(\mathbf r,t)= q(t)(V_{em}(\mathbf r)  + V_{d}(\mathbf r))$, where $V_{em}= E x$ is the potential of the uniform electric field $E$ near the surface of the MT; the time-dependent charge of the trailing head is $q(t)$, and $V_{d}$ is the docking potential.

The periodic docking potential we define as follows: $V_{d}(\mathbf r) = \sum_{m,n} V_{mn} (\mathbf r) $, where  $m,n=0,\pm 1,\pm 2, \dots$ and
\begin{align}
	V_{mn} =& -\frac{a_1 {\rm e}^{(z_0-r_{mn})/\kappa}}{\sqrt{(x-x_{mn})^2 + (y-y_{mn})^2+ z_0^2}}.
\end{align}
The pocket is located at the point $(x_{mn},y_{mn})$ of the docking site ($\beta$-tubulin subunit) belonging to $m$-th PF. The constant, $z_0$, denotes the minimum distance between the effective charge of the head and the surface of the MT. The parameter, $\kappa $, characterizes the Debye radius.

 In Fig. \ref{DP2a}, the docking potential landscape is depicted for three PFs ($m=-1, 0, 1$). The asymmetry in the docking potential is produced by the neighboring PFs. The kinesin moves along the central PF ($m=0$) in the positive direction of the $x$-axis. The trajectory of the left head is shown in blue, and the trajectory of the right head is shown in red. For the  given choice of parameters, the ratio of the maximum off-axis displacements for the right and left heads is $\approx 1$.
 
  \begin{figure}[tbh]
   \scalebox{0.3}{\includegraphics{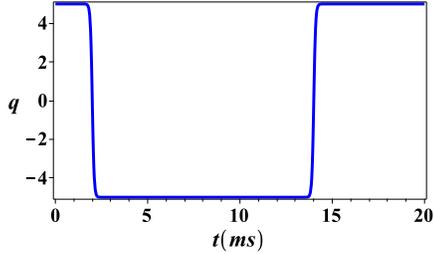}}
   \label{Ch}
 \caption{(Color online) Charge of the trailing head (in units of the electron charge, $ e$) as a function of time. Parameters: $\beta_1 = 10 \,\rm ms^{-1}$, $\beta_2 =10 \,\rm ms^{-1}$, $t_0 = 2 \,\rm ms$, and $\delta = 12 \,\rm ms$. 
 \label{DP2}}
  \end{figure}
   
   The function, $q(t)$, describes a redistribution of the total charge inside of the trailing kinesin head due to a conformation provided by the  ATP hydrolysis cycle: $q_0 \rightarrow -q_0 \rightarrow q_0$. At the beginning of the first step,  the trailing head has a negative charge.  At the end of the first step, the trailing head becomes positively charged and it attaches the $\beta$-tubulin.  
   
 We choose, $q(t)$, as a pulse with the shape given by,
\begin{align}
 	q(t) = &q_0 -\frac{q_0}{\tanh\Big(\displaystyle\frac{\delta}{2}\Big)}\big(\tanh(\beta_1(t - t_0))\nonumber \\
 	&- \tanh(\beta_2(t - t_0 - \delta)) ,
 \end{align}
 where $q_0$ is the charge of the kinesin head, and $t_0$  and $t_1= t_0+ \delta$ define  the times of unlocking (locking) of the docking sites. (See Fig. \ref{DP2}.)

It is known that the inter-head tension is required for normal kinesin motility \cite{Yildiz08}. We relate the external force,  $\mathbf F_{ex} =\mathbf F_{el}  + \mathbf F_{\tau}  $,  to the nonlinear elastic interaction between linked heads in the domain of the neck linker, and to the coiled  spring, located in the domain of neck coiled coil \cite{JH}. Here $\mathbf F_{el} $ stands for the elastic force in the neck linker, and $ \mathbf F_{\tau}  $ is the force caused by torque of the coiled spring \cite{Medina2009}.

To describe the elastic energy of the interaction between the heads  we use the finitely extensible nonlinear elastic (FENE) potential \cite{BAH,BK,KT},
\begin{align}
V_{el} = \left \{\begin{array}{ccl}
-\frac{k R_0^2}{2}\ln\bigg (1-  \frac{(l-l_1)^2}{R_o^2}\bigg),& |l-l_1| < R_0\\
\infty, & |l-l_1| > R_0
\end{array}
\right .
\end{align}
Here $l=|\mathbf r - \mathbf r_d|$ is the distance between the trailing  head and the locked head located at the point $\mathbf r_d$, $l_1$ being an equilibrium
distance, and $k$ is the stiffness of the string. The parameter, $R_0$, determines a maximum allowed separation: $l_1 - R_0 < l< l_1+ R_0$. For the elastic force acting on the trailing head, we obtain:  $ \mathbf F_{el}  =-\partial V_{el} /\partial \mathbf r$.
 
When the electric field drags the trailing head in the positive direction of the $x$-axis, the coiled spring is loaded. This results in the  additional force, $\mathbf F_{\tau}  $, acting on the head due the internal torque, $\tau_\alpha$, of the coiled spring. We assume that the potential energy stored in the coiled spring can be written as, $V_\tau=2\tau_\alpha N^2\sin^2 (\varphi/2N) $, where $\varphi$ is the turning angle, and $N$ is the number of the active coils. Thus, the minimum of the potential energy corresponds to,  $\varphi =0$. When the external force is applied to load the spring, the potential energy stored in spring is given by, $\Delta V_\tau=2\tau_\alpha N^2\sin^2 (\Delta \varphi/2N) $, where $\Delta \varphi $ is the deflection angle.

{\em Stepping dynamics.} -- We assume that the bounded (leading) head of the kinesin is initially located on the surface of the MT, at the  origin of the coordinates. The position of the tethered (tailing) is described by the radius vector, $\mathbf r = r( \cos \alpha,\sin \alpha,0)$. Assuming that the coiled spring is unloaded, we obtain, $\varphi = \pi - \alpha$.

Due to the asymmetry of the docking potential, when the rear head is released, the electric field pushes the head out of the pocket to the right, in the negative direction of the $y$-axis. Next, the uniform electric field of the MT drags the head in the positive direction of the $x$-axis. Thus, at the first step  the trailing ({\em right}) head is rotating in the counterclockwise direction. Its motion is described by Eqs. (\ref{EqL}) with,
\begin{align} 
	 \mathbf F_{\tau}   = -\Theta (t-t_0) {\tau_\alpha N}\sin\bigg( \frac{\pi - \alpha}{N}\bigg) \nabla \alpha,
\end{align}
where  $\Theta(t-t_0)$ is the Heaviside step function. When the right head  reaches the next docking site, it is locked, and, for a while, both heads are in the locked state. 
\begin{figure}[tbh]
    \scalebox{0.175}{\includegraphics{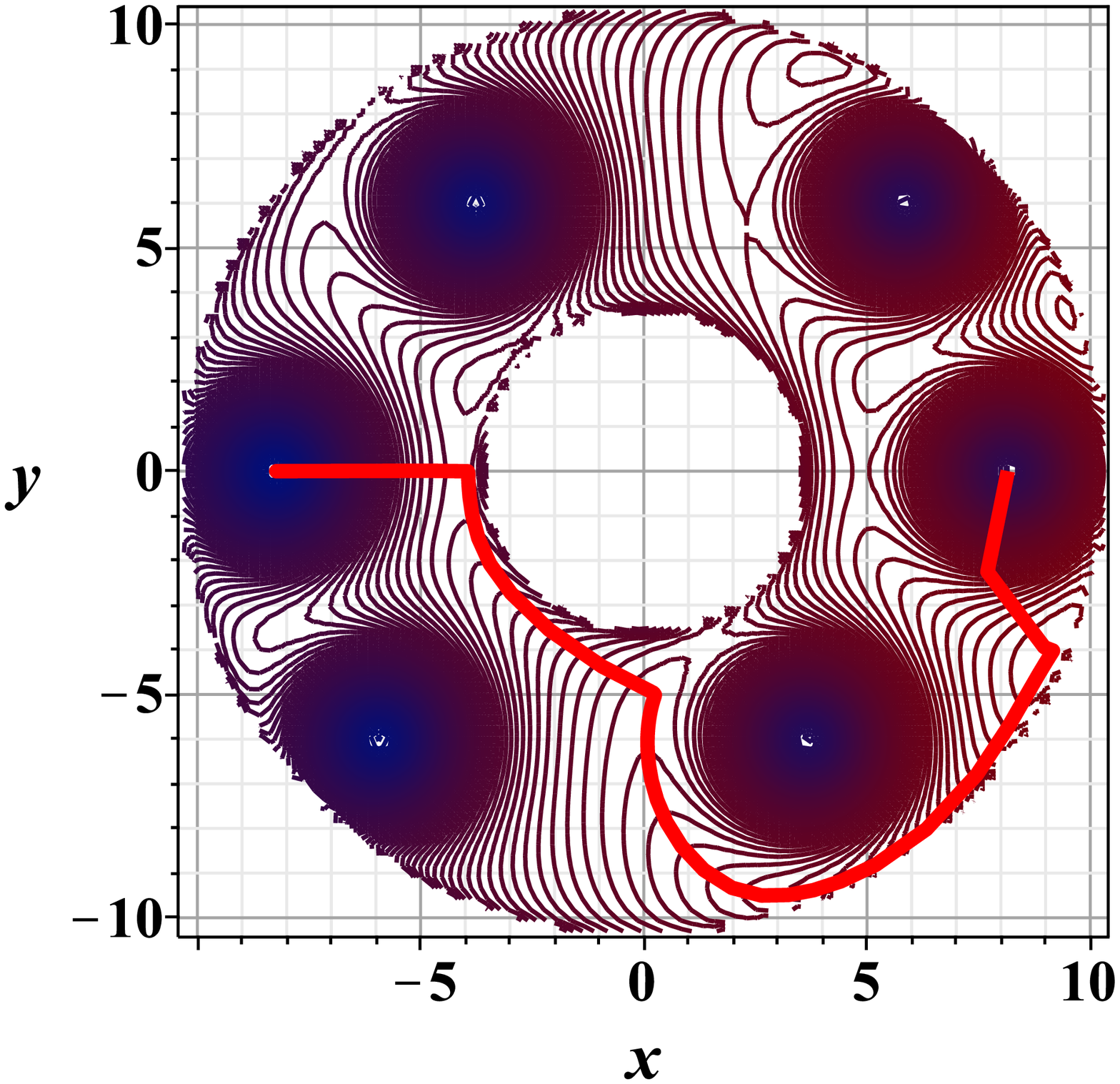}}
    \scalebox{0.175}{\includegraphics{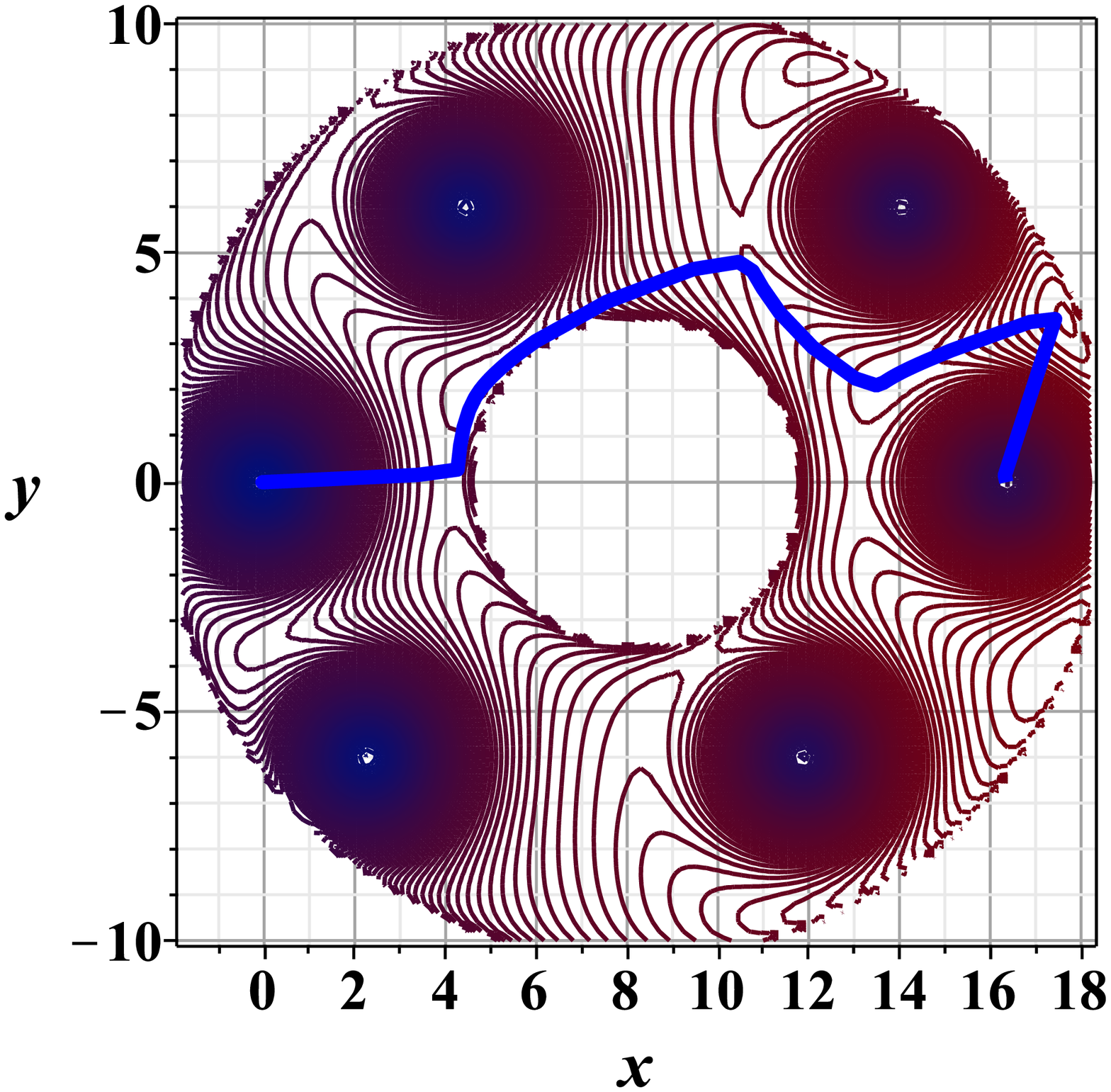}}
 \caption{(Color online) Total potential energy landscape for the unbounded heads.  Left panel: trajectory of the right trailing head (red). Right panel: trajectory of left trailing head (blue). 
 The stiffness of the neck linker is: $k = 2\, \rm pN /nm$, $\tau_\alpha = 0.5\, \rm pN\cdot nm \cdot rad^{-1}$.
 \label{CP1}}
  \end{figure}
\begin{figure}[tbh]
  \scalebox{0.325}{\includegraphics{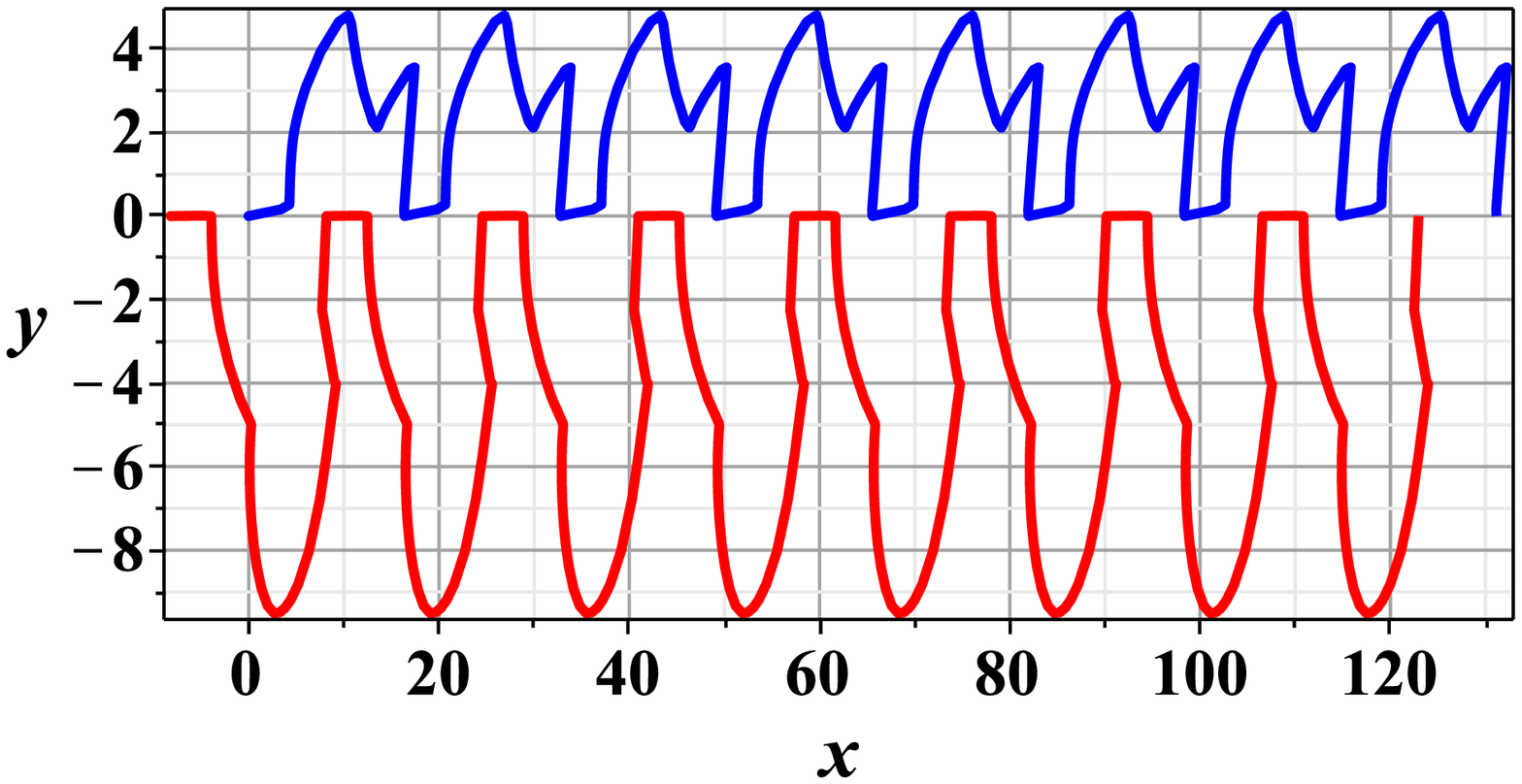}}
  (a)
    \scalebox{0.325}{\includegraphics{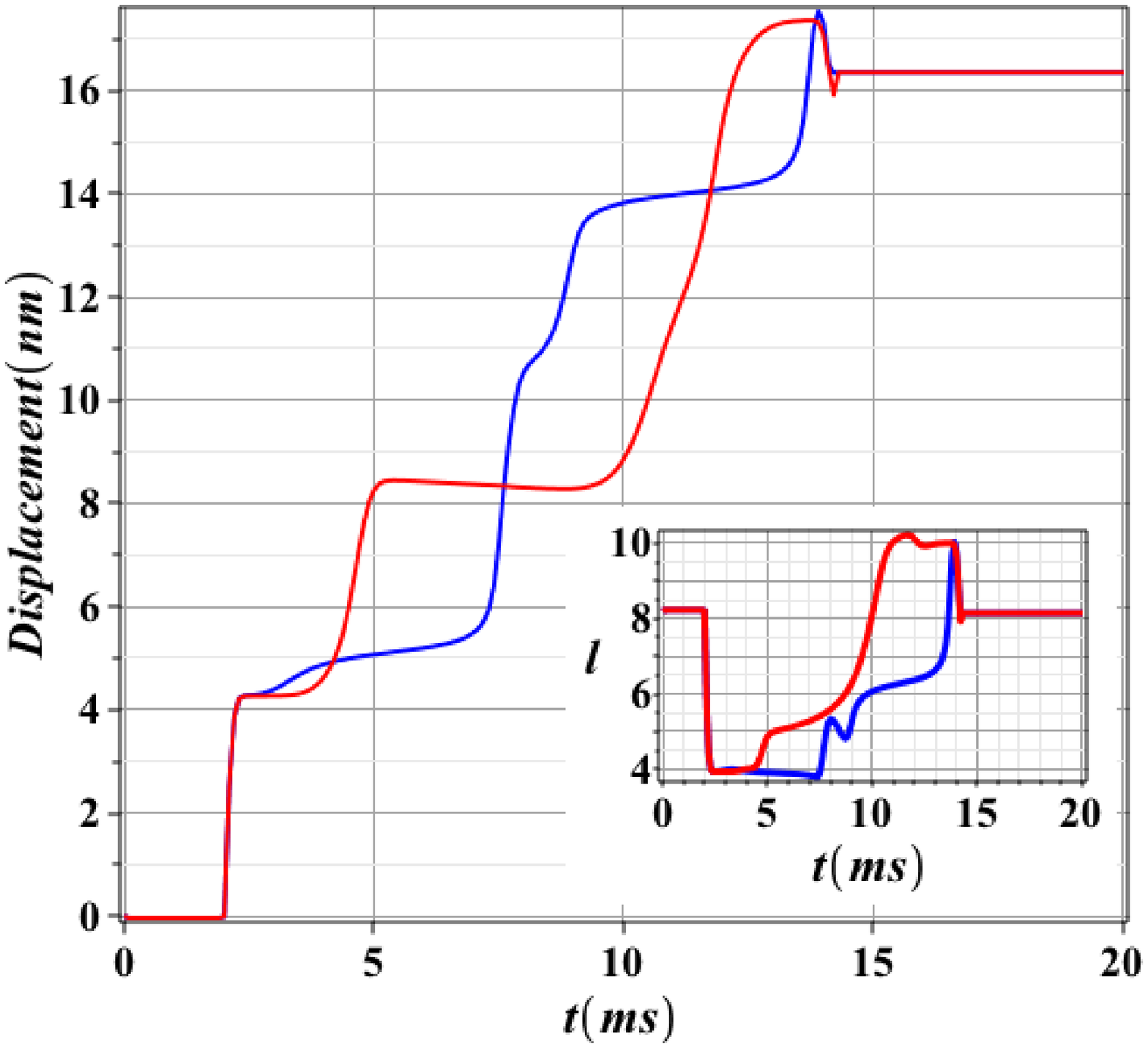}}
  (b)
 \caption{(Color online) (a) Kinesin steps: the average step size is about $16.4\, \rm nm$. (b) Displacement as a function of time. Inset: Dependence on time of the distance between heads, $l(t)$). Left trailing head (blue), right trailing head (red). 
  \label{KW}}
  \end{figure}
  
The second step begins when the formerly leading head (now the trailing head) is released. Since, during the first step the potential energy, $\Delta V_\tau=2\tau_\alpha N^2\sin^2 (\pi/2N) $, was stored in the coiled spring, the trailing head experiences now the torque generated by the coiled spring. The force, acting on the trailing head due to the torque of the coiled spring, can be written as,
\begin{align} 
	 \mathbf F_{\tau}   = \Theta (t-t_1) {\tau_\alpha N}\sin\bigg( \frac{2\pi - \alpha}{N}\bigg) \nabla \alpha,
\end{align}
where $t_1$ is the dwell time. Now the rotation of the head occurs in the clockwise direction. Below, we call this head  the {\em left head}. When the second step is completed, the kinesin is in the same state as before its first step: both heads are attached to the  neighboring $\beta$-tubulins, and the cycle is repeated.   

Note, that the direction of rotation of the left head depends on the torque and the conformation processes inside the kinesin.  If the coiled spring was not relaxed, or was not relaxed enough, the rotation of the trailing head will be performed in the clockwise direction. However, if during the stable state, with both heads being attached, the coiled spring was relaxed, the second step would occur in the counterclockwise rotation. Motion in which both heads of the kinesin rotate in the counterclockwise direction was observed in \cite{HRY}.  

\begin{figure}[tbh]
  \begin{center}
     \scalebox{0.275}{\includegraphics{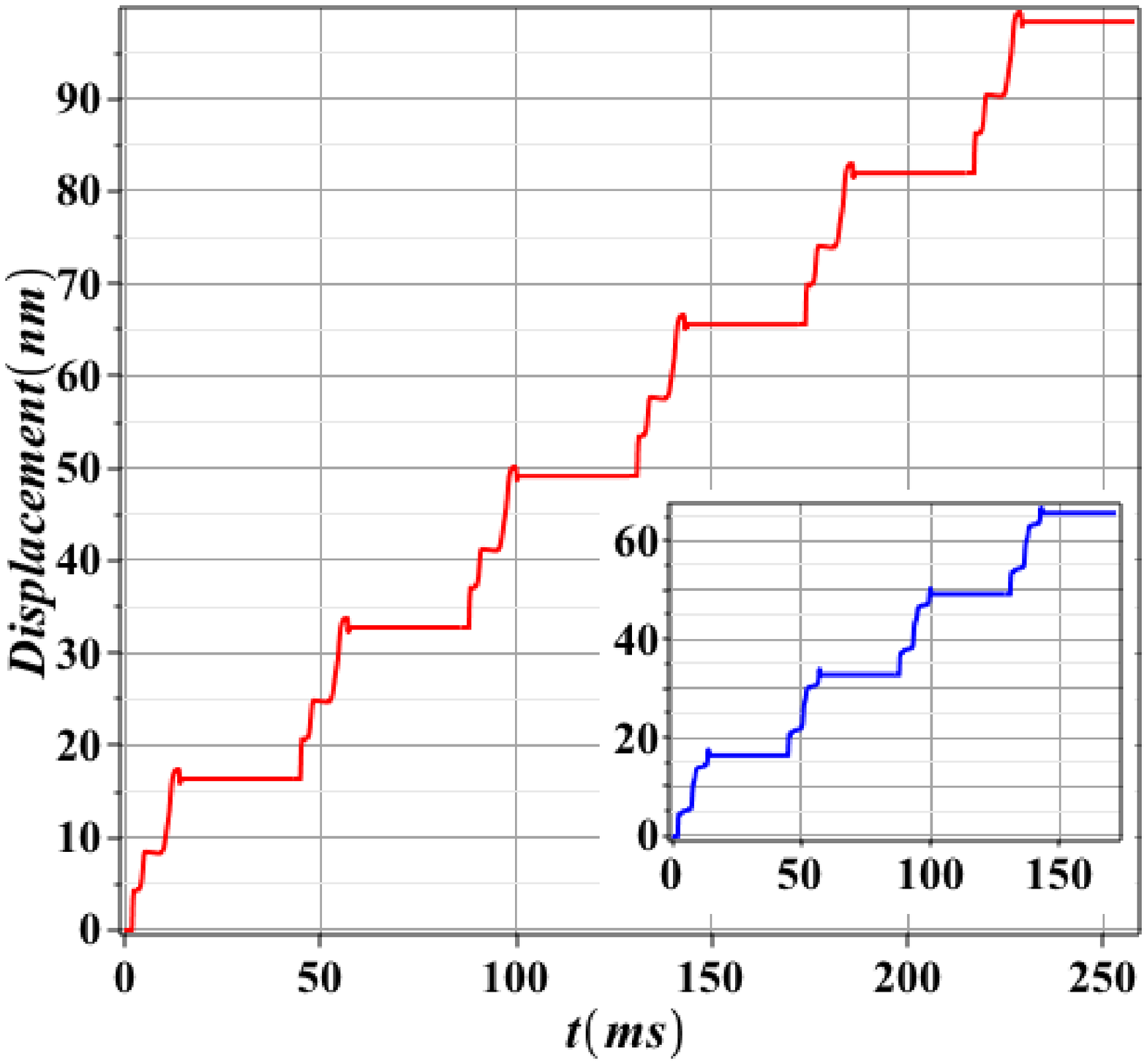}}
     (a)
  \scalebox{0.2}{\includegraphics{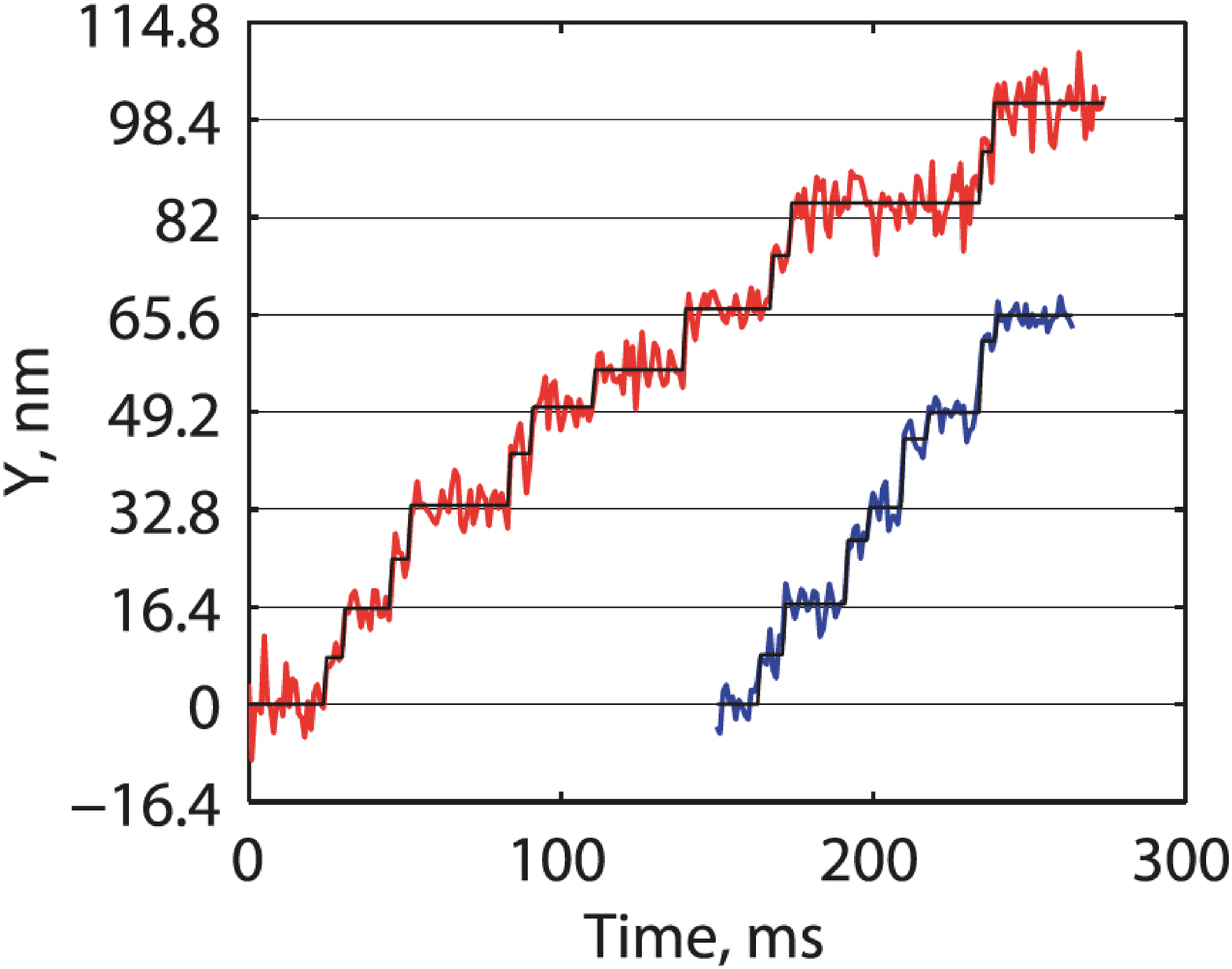}}
  (b)
 \end{center}
 \caption{(Color online)  The trajectories show the displacement of the right 
 trailing  head (red) and the left  trailing head (blue) along the 
 MT.  (a) Results of numerical simulations  demonstrate the occurrence of 
 substeps. (b) Experimental data taken from \cite{MDO}. Printed with 
 permission of PNAS.
 \label{KW2}}
  \end{figure}
\begin{figure}[tbh]
  \begin{center}
     \scalebox{0.165}{\includegraphics{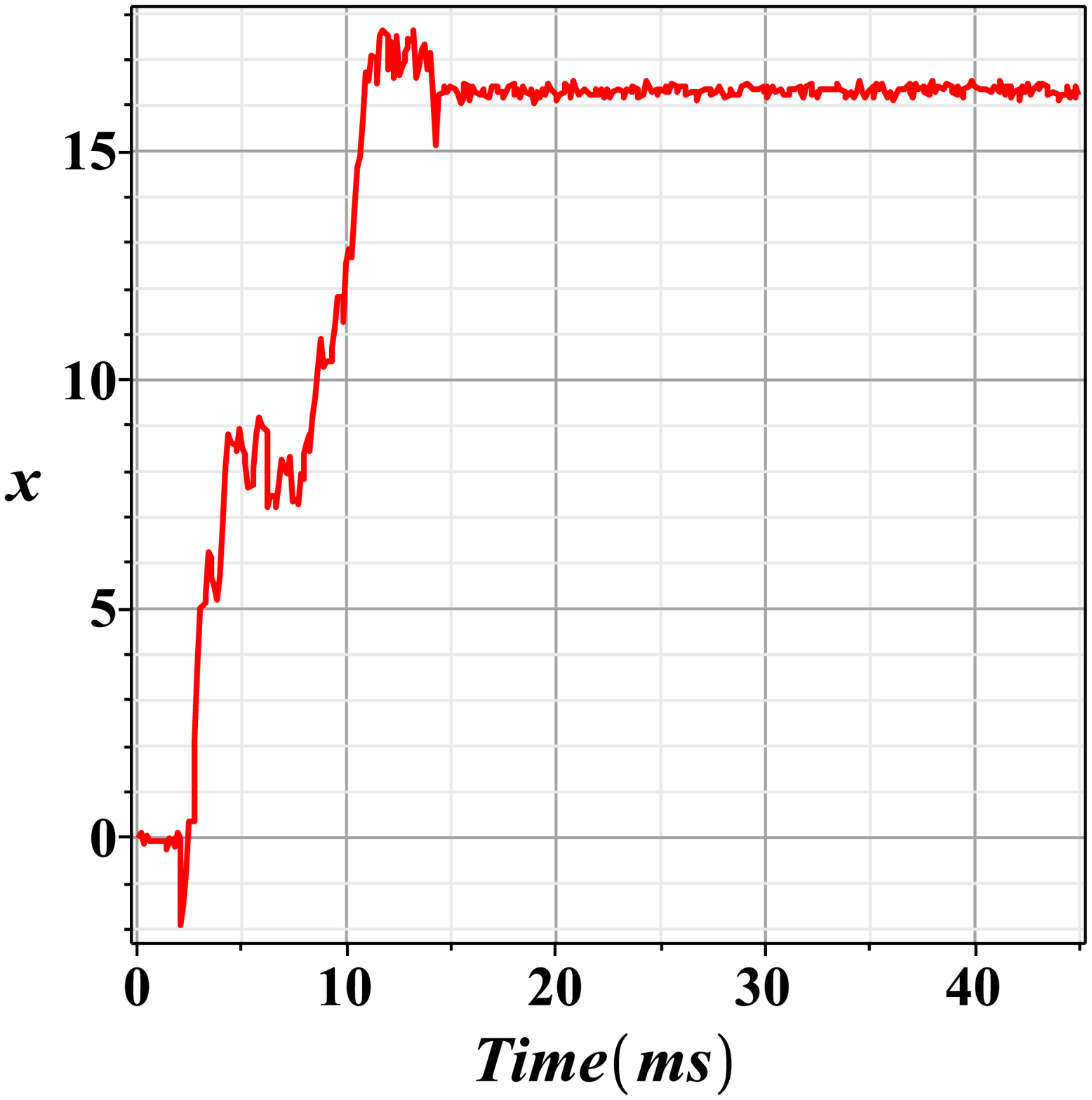}} 
     \scalebox{0.165}{\includegraphics{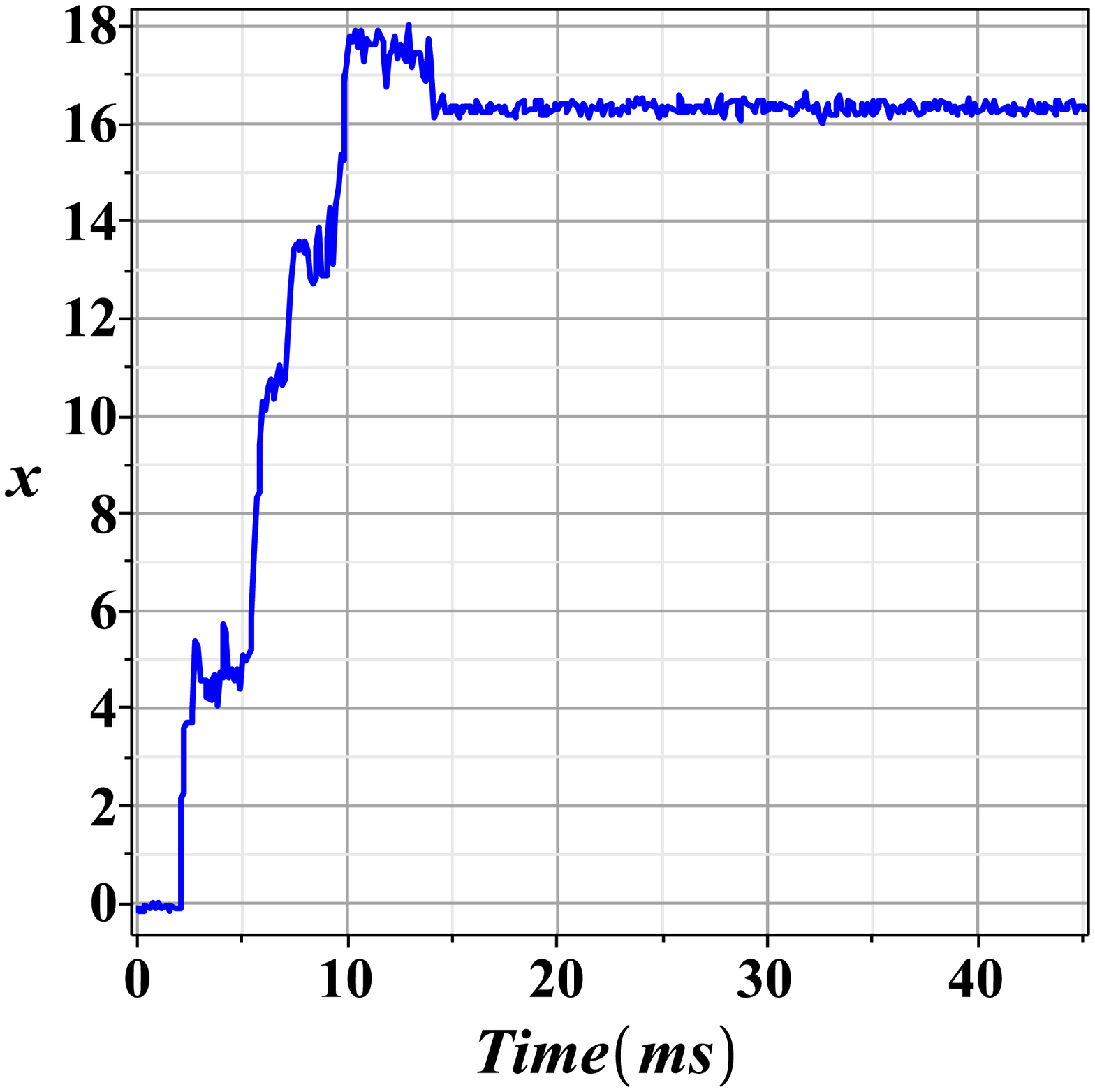}} 
     (a)
     \scalebox{0.165}{\includegraphics{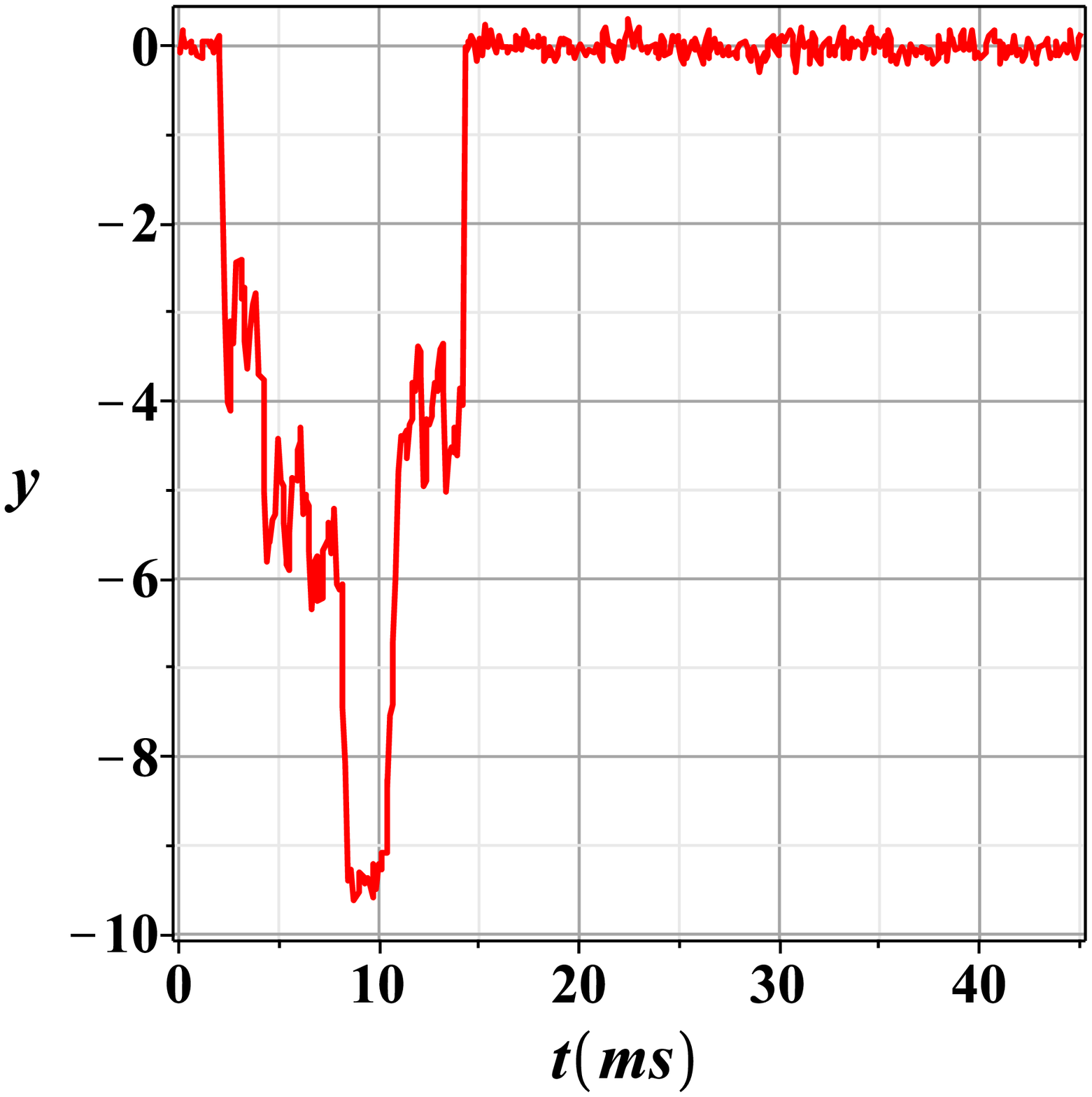}}
     \scalebox{0.165}{\includegraphics{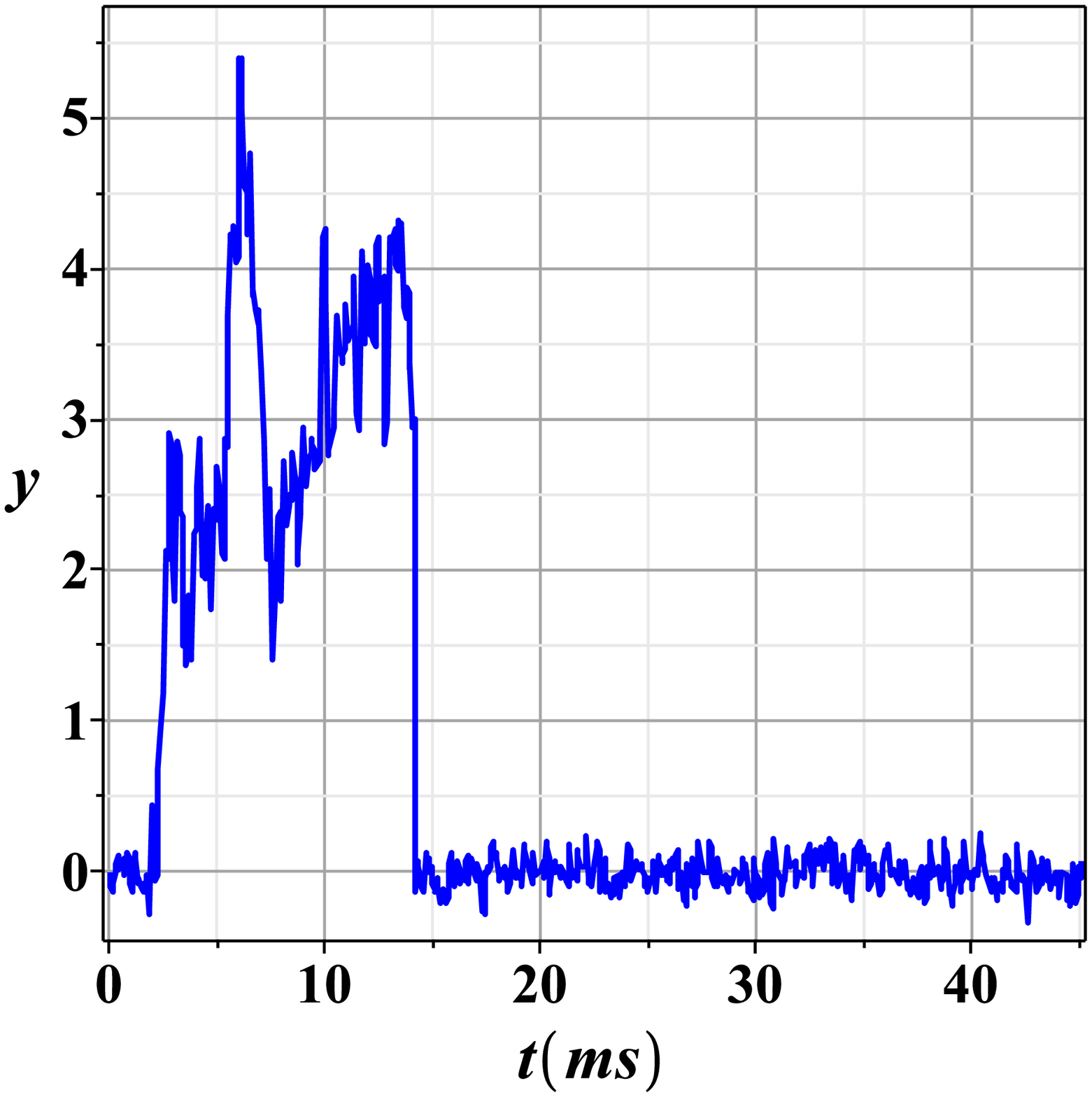}}
     (b)
 \end{center}
 \caption{(Color online)  Examples of trajectories  show the position of the  kinesin trailing head vs time: (a) position on the $x$-axis,  (b) position off the $x$-axis. Trajectory of the right (left) heads is shown by the red (blue) color. For each head, one can observe clear evidence of the substeps. The stiffness of the neck linker is: $k = 2\, \rm pN /nm$, $\tau_\alpha = 0.5\, \rm pN\cdot nm \cdot rad^{-1}$.
 \label{KW3}}
  \end{figure}
  
{\em Numerical results.} --  From the recent experimental data it follows that the average kinesin head displacement in the $x$-direction is: $ 2 l_0\approx 16.4\, \rm nm$ \cite{MDO,HRY}. Parameters used in our numerical simulations were chosen to obtain the best fit to the experimental data presented in \cite{MDO}. We chose: $N=2$; $\gamma= 2.75 \, \rm pN\cdot ms/ nm$; $V_0 = 150\, \rm pN\cdot nm$;  $\tau_\alpha = 0.5 \, \rm pN\cdot nm/rad$; $l_0 = 8.2 \, \rm  nm $; $l_1 = 7\, \rm nm$; $R_0 = 3.5 \, \rm  nm$; $k = 2\, \rm pN /nm$ \footnote{The results presented in Fig. \ref{DP2a} are obtained for  $k = 1.75\, \rm pN /nm$. }; $V_1 = 10^3 \, \rm pN\cdot nm^2$; $z_0=2 \, \rm  nm$; $l_1=7 \, \rm  nm$; $\kappa = 0.75\, \rm nm$ \footnote{Choice of $\kappa$ is based on the estimations in Ref. \cite{TIK}}. Here we set: $V_0 = q_0 l_0E$ and $V_1= q_0 a_1 $.

Our computation produces reasonable values of  the  electric field on the surface of the MT and the energy stored in the coiled spring after the first step: $E \sim 10^6 \, \rm V/m$ and $\Delta V_{\tau} \approx 1.57 \, \rm pN\cdot nm$. The average elastic force  is found to be $\langle F_{el} \rangle \approx 5 \, \rm pN$. This is in agreement with available data for the magnitude of the drag force \cite{HHG}.

In   Figs. \ref{CP1} -- \ref{KW2} the results of the numerical simulations, 
without environmental noise,  are presented.  As one can observe, the neck 
linker forces the trailing  head to move forward by $\approx 2\, \rm nm$. 
This is in a good agreement with the available experimental data on 
displacement of the connection point of head forward by  $\sim 2.7\, \rm 
nm$ \cite{Block2003}. The results obtained show clear evidence of substeps: 
one substep for the right head and two substeps for the left head. The 
appearance of substeps is due to the interaction with the (repulsive) docking 
potentials located at the sites on the neighboring PFs (Fig. \ref{CP1}). As 
shown in Figs. \ref{CP1} and \ref{KW}, the ratio of the maximum off-axis 
displacements for the right and left heads is $\approx 2$. This result agrees 
with experimental data obtained in \cite{MDO}.

 In Fig. \ref{KW3}, the results of solution of the Langevin stochastic equation (\ref{EqL}) are presented. To obtain the numerical solution, It${\rm \hat  o}$ calculus has been applied, with the diffusion parameter, $\sigma= \sqrt{2 \rm k_B T/\gamma}$, being $\sigma =1.74\, \rm ms\cdot nm^{-1}$. 
 
 Our results, presented in Fig.  \ref{KW} --  Fig. \ref{KW3}, show good agreement with the experimental data obtained in \cite{MDO}. (See Fig. \ref{KW2}b.) 
   
 \begin{acknowledgments}
 We are thankful to G.D. Doolen for useful comments. The work by G.P.B. was carried out under the auspices of the National Nuclear Security Administration of the U.S. Department of Energy at Los Alamos National Laboratory under Contract No. DE-AC52-06NA25396. A.I.N. and M.F.R. acknowledge the
support from the CONACyT. 
 \end{acknowledgments}


\begin{thebibliography}{100}

\expandafter\ifx\csname natexlab\endcsname\relax\def\natexlab#1{#1}\fi
\expandafter\ifx\csname bibnamefont\endcsname\relax
  \def\bibnamefont#1{#1}\fi
\expandafter\ifx\csname bibfnamefont\endcsname\relax
  \def\bibfnamefont#1{#1}\fi
\expandafter\ifx\csname citenamefont\endcsname\relax
  \def\citenamefont#1{#1}\fi
\expandafter\ifx\csname url\endcsname\relax
  \def\url#1{\texttt{#1}}\fi
\expandafter\ifx\csname urlprefix\endcsname\relax\def\urlprefix{URL }\fi
\providecommand{\bibinfo}[2]{#2}
\providecommand{\eprint}[2][]{\url{#2}}

\bibitem[{\citenamefont{Amos and Klug}(1974)}]{Amos}
\bibinfo{author}{\bibfnamefont{L.~A.} \bibnamefont{Amos}} 
\bibnamefont{and}
  \bibinfo{author}{\bibfnamefont{A.}~\bibnamefont{Klug}},
  \bibinfo{journal}{Journal of Cell Science} \textbf{\bibinfo{volume}{14}},
  \bibinfo{pages}{523} (\bibinfo{year}{1974}).

\bibitem[{\citenamefont{{ J. A. Tuszy\'nski, J. A. Brown, P. Hawrylak and P.
  Marcer}}(1998)}]{TBHM}
\bibinfo{author}{\bibnamefont{{ J. A. Tuszy\'nski, J. A. Brown, P. Hawrylak 
and
  P. Marcer}}}, \bibinfo{journal}{Phil. Trans. R. Soc. Lond. A}
  \textbf{\bibinfo{volume}{356}}, \bibinfo{pages}{1897} (\bibinfo{year}{1998}).

\bibitem[{\citenamefont{{ N. A. Baker, D. Sept, S. Joseph, M. J. Holst and J.
  A. McCammon}}(2001)}]{BSJ}
\bibinfo{author}{\bibnamefont{{ N. A. Baker, D. Sept, S. Joseph, M. J. Holst
  and J. A. McCammon}}}, \bibinfo{journal}{PNAS} 
  \textbf{\bibinfo{volume}{98}},
  \bibinfo{pages}{10037} (\bibinfo{year}{2001}).

\bibitem[{\citenamefont{{Nesterov} et~al.}(2016)\citenamefont{{Nesterov},
  {Ram{\'{\i}}rez}, {Berman}, and {Mavromatos}}}]{NRB}
\bibinfo{author}{\bibfnamefont{A.~I.} \bibnamefont{{Nesterov}}},
  \bibinfo{author}{\bibfnamefont{M.~F.} \bibnamefont{{Ram{\'{\i}}rez}}},
  \bibinfo{author}{\bibfnamefont{G.~P.} \bibnamefont{{Berman}}},
  \bibnamefont{and} \bibinfo{author}{\bibfnamefont{N.~E.}
  \bibnamefont{{Mavromatos}}}, \bibinfo{journal}{ArXiv e-prints}
  (\bibinfo{year}{2016}), \eprint{1604.05971}.

\bibitem[{\citenamefont{{ H. Bolterauer, J. A. Tuszynski2, and E.
  Unger}}(2005)}]{BTU}
\bibinfo{author}{\bibnamefont{{ H. Bolterauer, J. A. Tuszynski2, and E.
  Unger}}}, \bibinfo{journal}{Cell Biochem. Biophys.}
  \textbf{\bibinfo{volume}{42}}, \bibinfo{pages}{95 –}
  (\bibinfo{year}{2005}).

\bibitem[{\citenamefont{{J. M. Berg, J. L. Tymoczko, L. Stryer, and G. J.
  Gatto, Jr}}(2012)}]{BTS}
\bibinfo{author}{\bibnamefont{{J. M. Berg, J. L. Tymoczko, L. Stryer, and G. J.
  Gatto, Jr}}}, \emph{\bibinfo{title}{Biochemistry}} (\bibinfo{publisher}{W. H.
  Freeman and Company, N Y}, \bibinfo{year}{2012}).

\bibitem[{\citenamefont{Boehm et~al.}(1997)\citenamefont{Boehm, 
Steinmetzer,
  Daniel, W.Vater, Baum, and Unger}}]{BSV}
\bibinfo{author}{\bibfnamefont{K.~J.} \bibnamefont{Boehm}},
  \bibinfo{author}{\bibfnamefont{P.}~\bibnamefont{Steinmetzer}},
  \bibinfo{author}{\bibfnamefont{A.}~\bibnamefont{Daniel}},
  \bibinfo{author}{\bibnamefont{W.Vater}},
  \bibinfo{author}{\bibfnamefont{M.}~\bibnamefont{Baum}}, 
  \bibnamefont{and}
  \bibinfo{author}{\bibfnamefont{E.}~\bibnamefont{Unger}},
  \bibinfo{journal}{Cell Motil. Cytoskeleton} \textbf{\bibinfo{volume}{37}},
  \bibinfo{pages}{226 – 231} (\bibinfo{year}{1997}).

\bibitem[{\citenamefont{Vale et~al.}(1985)\citenamefont{Vale, Reese, and
  Sheetz}}]{VRS}
\bibinfo{author}{\bibfnamefont{R.}~\bibnamefont{Vale}},
  \bibinfo{author}{\bibfnamefont{T.}~\bibnamefont{Reese}}, 
  \bibnamefont{and}
  \bibinfo{author}{\bibfnamefont{M.~P.} \bibnamefont{Sheetz}},
  \bibinfo{journal}{Cell} \textbf{\bibinfo{volume}{42}}, \bibinfo{pages}{39 }
  (\bibinfo{year}{1985}).

\bibitem[{\citenamefont{von Massow et~al.}(1989)\citenamefont{von 
Massow,
  Mandelkow, and Mandelkow}}]{MMM}
\bibinfo{author}{\bibfnamefont{A.}~\bibnamefont{von Massow}},
  \bibinfo{author}{\bibfnamefont{E.}~\bibnamefont{Mandelkow}},
  \bibnamefont{and}
  \bibinfo{author}{\bibfnamefont{E.}~\bibnamefont{Mandelkow}},
  \bibinfo{journal}{Cell Motil Cytoskeleton} \textbf{\bibinfo{volume}{92}},
  \bibinfo{pages}{562 } (\bibinfo{year}{1989}).

\bibitem[{\citenamefont{Howard et~al.}(1989)\citenamefont{Howard, 
Hudspeth, and
  Vale}}]{HVR}
\bibinfo{author}{\bibfnamefont{J.}~\bibnamefont{Howard}},
  \bibinfo{author}{\bibfnamefont{A.}~\bibnamefont{Hudspeth}}, 
  \bibnamefont{and}
  \bibinfo{author}{\bibfnamefont{R.}~\bibnamefont{Vale}},
  \bibinfo{journal}{Nature} \textbf{\bibinfo{volume}{342}},
  \bibinfo{pages}{154} (\bibinfo{year}{1989}).

\bibitem[{\citenamefont{Hancock and Howard}(1999)}]{HHJ}
\bibinfo{author}{\bibfnamefont{W.~O.} \bibnamefont{Hancock}} 
\bibnamefont{and}
  \bibinfo{author}{\bibfnamefont{J.}~\bibnamefont{Howard}},
  \bibinfo{journal}{PNAS} \textbf{\bibinfo{volume}{96}},
  \bibinfo{pages}{13147–} (\bibinfo{year}{1999}).

\bibitem[{\citenamefont{Hackney}(1994)}]{HD}
\bibinfo{author}{\bibfnamefont{D.}~\bibnamefont{Hackney}},
  \bibinfo{journal}{PNAS} \textbf{\bibinfo{volume}{13}},
  \bibinfo{pages}{6865–} (\bibinfo{year}{1994}).

\bibitem[{\citenamefont{Schief and Howard}(2001)}]{SHJ}
\bibinfo{author}{\bibfnamefont{W.}~\bibnamefont{Schief}} \bibnamefont{and}
  \bibinfo{author}{\bibfnamefont{J.}~\bibnamefont{Howard}},
  \bibinfo{journal}{Curr Opin Cell Biol} \textbf{\bibinfo{volume}{13}},
  \bibinfo{pages}{19} (\bibinfo{year}{2001}).

\bibitem[{\citenamefont{Asbury et~al.}(2003)\citenamefont{Asbury, Fehr, and
  Block}}]{AFB}
\bibinfo{author}{\bibfnamefont{C.}~\bibnamefont{Asbury}},
  \bibinfo{author}{\bibfnamefont{A.}~\bibnamefont{Fehr}}, \bibnamefont{and}
  \bibinfo{author}{\bibfnamefont{S.}~\bibnamefont{Block}},
  \bibinfo{journal}{Science} \textbf{\bibinfo{volume}{302}},
  \bibinfo{pages}{2130–} (\bibinfo{year}{2003}).

\bibitem[{\citenamefont{Yildiz et~al.}(2004)\citenamefont{Yildiz, Tomishige,
  Vale, and Selvin}}]{Yild04}
\bibinfo{author}{\bibfnamefont{A.}~\bibnamefont{Yildiz}},
  \bibinfo{author}{\bibfnamefont{M.}~\bibnamefont{Tomishige}},
  \bibinfo{author}{\bibfnamefont{R.~D.} \bibnamefont{Vale}}, 
  \bibnamefont{and}
  \bibinfo{author}{\bibfnamefont{P.~R.} \bibnamefont{Selvin}},
  \bibinfo{journal}{Science} \textbf{\bibinfo{volume}{303}},
  \bibinfo{pages}{676} (\bibinfo{year}{2004}).

\bibitem[{\citenamefont{Schaap et~al.}(2011)\citenamefont{Schaap, Carrasco,
  de~Pablo, and Schmidt}}]{Schaap2011}
\bibinfo{author}{\bibfnamefont{I.~A.} \bibnamefont{Schaap}},
  \bibinfo{author}{\bibfnamefont{C.}~\bibnamefont{Carrasco}},
  \bibinfo{author}{\bibfnamefont{P.}~\bibnamefont{de~Pablo}}, 
  \bibnamefont{and}
  \bibinfo{author}{\bibfnamefont{C.~F.} \bibnamefont{Schmidt}},
  \bibinfo{journal}{Biophys J} \textbf{\bibinfo{volume}{100}},
  \bibinfo{pages}{2450} (\bibinfo{year}{2011}).

\bibitem[{\citenamefont{Mickolajczyk 
et~al.}(2015)\citenamefont{Mickolajczyk,
  Deffenbaugh, Ortega~Arroyo, Andrecka, Kukura, and Hancock}}]{MDO}
\bibinfo{author}{\bibfnamefont{K.~J.} \bibnamefont{Mickolajczyk}},
  \bibinfo{author}{\bibfnamefont{N.~C.} \bibnamefont{Deffenbaugh}},
  \bibinfo{author}{\bibfnamefont{J.}~\bibnamefont{Ortega~Arroyo}},
  \bibinfo{author}{\bibfnamefont{J.}~\bibnamefont{Andrecka}},
  \bibinfo{author}{\bibfnamefont{P.}~\bibnamefont{Kukura}}, 
  \bibnamefont{and}
  \bibinfo{author}{\bibfnamefont{W.~O.} \bibnamefont{Hancock}},
  \bibinfo{journal}{PNAS} \textbf{\bibinfo{volume}{112}},
  \bibinfo{pages}{E7186} (\bibinfo{year}{2015}).

\bibitem[{\citenamefont{Isojima et~al.}(2016)\citenamefont{Isojima, Iino,
  Niitani, Noji, and Tomishige}}]{HRY}
\bibinfo{author}{\bibfnamefont{H.}~\bibnamefont{Isojima}},
  \bibinfo{author}{\bibfnamefont{R.}~\bibnamefont{Iino}},
  \bibinfo{author}{\bibfnamefont{Y.}~\bibnamefont{Niitani}},
  \bibinfo{author}{\bibfnamefont{H.}~\bibnamefont{Noji}}, \bibnamefont{and}
  \bibinfo{author}{\bibfnamefont{M.}~\bibnamefont{Tomishige}},
  \bibinfo{journal}{Nat. Chem. Biol.} \textbf{\bibinfo{volume}{12}},
  \bibinfo{pages}{290–297} (\bibinfo{year}{2016}).

\bibitem[{\citenamefont{Block}(2007)}]{Block2007}
\bibinfo{author}{\bibfnamefont{S.~M.} \bibnamefont{Block}},
  \bibinfo{journal}{Biophys. J.} \textbf{\bibinfo{volume}{92}},
  \bibinfo{pages}{2986} (\bibinfo{year}{2007}).

\bibitem[{\citenamefont{Thomas et~al.}(2002)\citenamefont{Thomas, 
Imafuku,
  Kamiya, and Tawada}}]{TIK}
\bibinfo{author}{\bibfnamefont{N.}~\bibnamefont{Thomas}},
  \bibinfo{author}{\bibfnamefont{Y.}~\bibnamefont{Imafuku}},
  \bibinfo{author}{\bibfnamefont{T.}~\bibnamefont{Kamiya}}, 
  \bibnamefont{and}
  \bibinfo{author}{\bibfnamefont{K.}~\bibnamefont{Tawada}},
  \bibinfo{journal}{Proc. Roy. Soc. B} \textbf{\bibinfo{volume}{269}},
  \bibinfo{pages}{2363 } (\bibinfo{year}{2002}).

\bibitem[{\citenamefont{Shao and Gao}(2006)}]{SQG}
\bibinfo{author}{\bibfnamefont{Q.}~\bibnamefont{Shao}} \bibnamefont{and}
  \bibinfo{author}{\bibfnamefont{Y.~Q.} \bibnamefont{Gao}},
  \bibinfo{journal}{PNAS} \textbf{\bibinfo{volume}{103}}, 
  \bibinfo{pages}{8072}
  (\bibinfo{year}{2006}).

\bibitem[{\citenamefont{Carter and Cross}(2005)}]{CCA}
\bibinfo{author}{\bibfnamefont{N.}~\bibnamefont{Carter}} \bibnamefont{and}
  \bibinfo{author}{\bibfnamefont{R.}~\bibnamefont{Cross}},
  \bibinfo{journal}{Nature} \textbf{\bibinfo{volume}{435}},
  \bibinfo{pages}{308} (\bibinfo{year}{2005}).

\bibitem[{\citenamefont{Bryant et~al.}(2003)\citenamefont{Bryant, Stone, 
Gore,
  Smith, Cozzarelli, and Bustamante}}]{BSG}
\bibinfo{author}{\bibfnamefont{Z.}~\bibnamefont{Bryant}},
  \bibinfo{author}{\bibfnamefont{M.~D.} \bibnamefont{Stone}},
  \bibinfo{author}{\bibfnamefont{J.}~\bibnamefont{Gore}},
  \bibinfo{author}{\bibfnamefont{S.~B.} \bibnamefont{Smith}},
  \bibinfo{author}{\bibfnamefont{N.~R.} \bibnamefont{Cozzarelli}},
  \bibnamefont{and}
  \bibinfo{author}{\bibfnamefont{C.}~\bibnamefont{Bustamante}},
  \bibinfo{journal}{Nature} \textbf{\bibinfo{volume}{424}},
  \bibinfo{pages}{338} (\bibinfo{year}{2003}).

\bibitem[{\citenamefont{Woehlke et~al.}(1997)\citenamefont{Woehlke, Ruby, 
Hart,
  Ly, Hom-Booher, and Vale}}]{Woehlke97}
\bibinfo{author}{\bibfnamefont{G.}~\bibnamefont{Woehlke}},
  \bibinfo{author}{\bibfnamefont{A.~K.} \bibnamefont{Ruby}},
  \bibinfo{author}{\bibfnamefont{C.~L.} \bibnamefont{Hart}},
  \bibinfo{author}{\bibfnamefont{B.}~\bibnamefont{Ly}},
  \bibinfo{author}{\bibfnamefont{N.}~\bibnamefont{Hom-Booher}},
  \bibnamefont{and} \bibinfo{author}{\bibfnamefont{R.~D.} 
  \bibnamefont{Vale}},
  \bibinfo{journal}{Cell} \textbf{\bibinfo{volume}{90}}, \bibinfo{pages}{207}
  (\bibinfo{year}{1997}).

\bibitem[{\citenamefont{Grant et~al.}(2011)\citenamefont{Grant, 
M.~Gheorghe,
  Zheng, Alonso, Huber, Dlugosz, McCammon, and Cross}}]{GGZ}
\bibinfo{author}{\bibfnamefont{B.~J.} \bibnamefont{Grant}},
  \bibinfo{author}{\bibfnamefont{D.}~\bibnamefont{M.~Gheorghe}},
  \bibinfo{author}{\bibfnamefont{W.}~\bibnamefont{Zheng}},
  \bibinfo{author}{\bibfnamefont{M.}~\bibnamefont{Alonso}},
  \bibinfo{author}{\bibfnamefont{G.}~\bibnamefont{Huber}},
  \bibinfo{author}{\bibfnamefont{M.}~\bibnamefont{Dlugosz}},
  \bibinfo{author}{\bibfnamefont{J.~A.} \bibnamefont{McCammon}},
  \bibnamefont{and} \bibinfo{author}{\bibfnamefont{R.~A.} 
  \bibnamefont{Cross}},
  \bibinfo{journal}{PLoS Biol} \textbf{\bibinfo{volume}{9}},
  \bibinfo{pages}{e1001207} (\bibinfo{year}{2011}).

\bibitem[{\citenamefont{Andrews et~al.}(1993)\citenamefont{Andrews, 
Gallant,
  Leapman, Schnapp, and Reese}}]{AGL}
\bibinfo{author}{\bibfnamefont{S.}~\bibnamefont{Andrews}},
  \bibinfo{author}{\bibfnamefont{P.}~\bibnamefont{Gallant}},
  \bibinfo{author}{\bibfnamefont{R.}~\bibnamefont{Leapman}},
  \bibinfo{author}{\bibfnamefont{B.}~\bibnamefont{Schnapp}}, 
  \bibnamefont{and}
  \bibinfo{author}{\bibfnamefont{T.}~\bibnamefont{Reese}},
  \bibinfo{journal}{PNAS} \textbf{\bibinfo{volume}{90}},
  \bibinfo{pages}{6503–} (\bibinfo{year}{1993}).

\bibitem[{\citenamefont{Svoboda and Block}(1994)}]{SBS}
\bibinfo{author}{\bibfnamefont{K.}~\bibnamefont{Svoboda}} 
\bibnamefont{and}
  \bibinfo{author}{\bibfnamefont{S.~M.} \bibnamefont{Block}},
  \bibinfo{journal}{Cell} \textbf{\bibinfo{volume}{77}}, \bibinfo{pages}{773}
  (\bibinfo{year}{1994}).

\bibitem[{\citenamefont{Hendricks and Erika L. F.~Holzbaur}(2012)}]{HHG}
\bibinfo{author}{\bibfnamefont{A.~G.} \bibnamefont{Hendricks}}
  \bibnamefont{and} \bibinfo{author}{\bibfnamefont{a.~E.~G.} 
  \bibnamefont{Erika
  L. F.~Holzbaur}}, \bibinfo{journal}{PNAS} \textbf{\bibinfo{volume}{109}},
  \bibinfo{pages}{18447} (\bibinfo{year}{2012}).

\bibitem[{\citenamefont{Leidel et~al.}(2012)\citenamefont{Leidel, Longoria,
  Gutierrez, and Shubeita}}]{LLG}
\bibinfo{author}{\bibfnamefont{C.}~\bibnamefont{Leidel}},
  \bibinfo{author}{\bibfnamefont{R.}~\bibnamefont{Longoria}},
  \bibinfo{author}{\bibfnamefont{F.}~\bibnamefont{Gutierrez}},
  \bibnamefont{and} 
  \bibinfo{author}{\bibfnamefont{G.}~\bibnamefont{Shubeita}},
  \bibinfo{journal}{Biophys. J.} \textbf{\bibinfo{volume}{103}},
  \bibinfo{pages}{18447} (\bibinfo{year}{2012}).

\bibitem[{\citenamefont{Yildiz et~al.}(2008)\citenamefont{Yildiz, Tomishige,
  Gennerich, and Vale}}]{Yildiz08}
\bibinfo{author}{\bibfnamefont{A.}~\bibnamefont{Yildiz}},
  \bibinfo{author}{\bibfnamefont{M.}~\bibnamefont{Tomishige}},
  \bibinfo{author}{\bibfnamefont{A.}~\bibnamefont{Gennerich}},
  \bibnamefont{and} \bibinfo{author}{\bibfnamefont{R.~D.} 
  \bibnamefont{Vale}},
  \bibinfo{journal}{Cell} \textbf{\bibinfo{volume}{134}}, \bibinfo{pages}{1030}
  (\bibinfo{year}{2008}).

\bibitem[{\citenamefont{Jeppesen and Hoerber}(2012)}]{JH}
\bibinfo{author}{\bibfnamefont{G.~M.} \bibnamefont{Jeppesen}} 
\bibnamefont{and}
  \bibinfo{author}{\bibfnamefont{J.~H.} \bibnamefont{Hoerber}},
  \bibinfo{journal}{Trans. Biochem. Soc.} \textbf{\bibinfo{volume}{40}},
  \bibinfo{pages}{438 } (\bibinfo{year}{2012}).

\bibitem[{\citenamefont{Guti{\'e}rrez-Medina
  et~al.}(2009)\citenamefont{Guti{\'e}rrez-Medina, Fehr, and
  Block}}]{Medina2009}
\bibinfo{author}{\bibfnamefont{B.}~\bibnamefont{Guti{\'e}rrez-Medina}},
  \bibinfo{author}{\bibfnamefont{A.~N.} \bibnamefont{Fehr}}, 
  \bibnamefont{and}
  \bibinfo{author}{\bibfnamefont{S.~M.} \bibnamefont{Block}},
  \bibinfo{journal}{PNAS} \textbf{\bibinfo{volume}{106}},
  \bibinfo{pages}{17007} (\bibinfo{year}{2009}).

\bibitem[{\citenamefont{Bird et~al.}(1991)\citenamefont{Bird, Armstrong, and
  Hassager}}]{BAH}
\bibinfo{author}{\bibfnamefont{R.}~\bibnamefont{Bird}},
  \bibinfo{author}{\bibfnamefont{R.}~\bibnamefont{Armstrong}},
  \bibnamefont{and} 
  \bibinfo{author}{\bibfnamefont{O.}~\bibnamefont{Hassager}},
  \emph{\bibinfo{title}{Dynamics of Polymeric Liquids V1,2}}
  (\bibinfo{publisher}{Wiley}, \bibinfo{year}{1991}).

\bibitem[{\citenamefont{{K. Binder (Ed.)}}(1995)}]{BK}
\bibinfo{author}{\bibnamefont{{K. Binder (Ed.)}}}, \emph{\bibinfo{title}{Monte
  Carlo and Molecular Dynamics Simulations in Polymer Science}}
  (\bibinfo{publisher}{Oxford}, \bibinfo{year}{1995}).

\bibitem[{\citenamefont{Kawakatsu}(2004)}]{KT}
\bibinfo{author}{\bibfnamefont{T.}~\bibnamefont{Kawakatsu}},
  \emph{\bibinfo{title}{Statistical Physics of Polymers}}
  (\bibinfo{publisher}{Springer}, \bibinfo{year}{2004}).

\bibitem[{\citenamefont{Block et~al.}(2003)\citenamefont{Block, Asbury,
  Shaevitz, and Lang}}]{Block2003}
\bibinfo{author}{\bibfnamefont{S.~M.} \bibnamefont{Block}},
  \bibinfo{author}{\bibfnamefont{C.~L.} \bibnamefont{Asbury}},
  \bibinfo{author}{\bibfnamefont{J.~W.} \bibnamefont{Shaevitz}},
  \bibnamefont{and} \bibinfo{author}{\bibfnamefont{M.~J.} 
  \bibnamefont{Lang}},
  \bibinfo{journal}{PNAS} \textbf{\bibinfo{volume}{100}}, 
  \bibinfo{pages}{2351}
  (\bibinfo{year}{2003}).

\end{thebibliography}
\end{document}